\documentclass{sigchi}
\usepackage[utf8]{inputenc}
\usepackage[english]{babel}
\usepackage{amsmath,amsfonts,amssymb,float,bbm}
\usepackage{balance}  % to better equalize the last page
\usepackage{graphicx} 
\usepackage{txfonts}
\usepackage{times}    % comment if you want LaTeX's default font
\usepackage[pdftex]{hyperref}
\usepackage{color}
\usepackage{textcomp}
\usepackage{booktabs}
\usepackage{ccicons}
\usepackage{todonotes}
\usepackage{url}

\toappear{}

\newcommand{\noop}[1]{} % Empty op

\makeatletter
\def\url@leostyle{%
  \@ifundefined{selectfont}{\def\UrlFont{\sf}}{\def\UrlFont{\small\bf\ttfamily}}}
\makeatother
\def\pprw{8.5in}
\def\pprh{11in}

\definecolor{linkColor}{RGB}{6,125,233}

\graphicspath{ {../Figuras/} }

\begin{document}
\title{Tweeting Over The Border: An Empirical Study of Transnational Migration in San Diego and Tijuana}

% \numberofauthors{1}
% \author{%
%   \alignauthor{[Hidden for blind review]}
% }

\numberofauthors{3}
\author{%
  \alignauthor{Víctor R. Martínez\\
    \affaddr{Instituto Tecnológico Autónomo de México}\\
    \affaddr{Río Hondo No.1, Mexico City, Mexico}\\
    \email{victor.martinez@itam.mx}}\\
  \alignauthor{Antonio Mancilla\\
    \affaddr{Instituto Tecnológico Autónomo de México}\\
    \affaddr{Río Hondo No.1, Mexico City, Mexico}\\
    \email{jose.mancilla@itam.mx}}\\
  \alignauthor{Víctor M. González\\
    \affaddr{Dept. of Computer Science \\ Instituto Tecnológico Autónomo de México}\\
    \affaddr{Río Hondo No.1, Mexico City, Mexico}\\
    \email{victor.gonzalez@itam.mx}}\\
}

\maketitle

\begin{abstract}
Sociological studies on transnational migration are often based on surveys or interviews, an expensive and time-consuming approach. On the other hand, the pervasiveness of mobile phones and location-aware social networks has introduced new ways to understand human mobility patterns at a national or global scale. In this work, we leverage geo-located information obtained from Twitter as to understand transnational migration patterns between two border cities (San Diego, USA and Tijuana, Mexico). We obtained 10.9M geo-located tweets from December 2013 to January 2015. Our method infers human mobility by inspecting tweet submissions and user’s home locations. Our results depict a trans-national community structure that exhibits the formation of a functional metropolitan area that physically transcends international borders. These results show the potential for re-analyzing sociology phenomena from a technology-based empirical perspective.
\end{abstract}


\section{Introduction}
Historically, international borders have denoted a clash of cultures, races, economies and governments \cite{Sparrow2001}. Toward the end of the twentieth century, the emergence of globalization and planet-scale communications have reduced the function of borders from a position as trade barriers to deterrence of human migrations. This is especially true in the case of U.S. -- Mexico border, the ninth longest frontier in the world. With over a million people crossing daily, this international border is considered the busiest in the world, as well as one of the most contrasting frontiers \cite{Lange1999a,romero2008hyperborder}. 

The largest U.S. -- Mexico border metropolitan region is the San Diego -- Tijuana region. It is at the most southwestern point of the U.S. and the most northwestern point of Mexico. Currently, it has a combined population of over 4.0 million people, anticipated to grow to over 5.5 million by 2020 \cite{Peach1999,sandag05}. San Diego and Tijuana present a unique relationship because of their extreme income difference (with a ratio of 6.4 to 1) and marked economical inequality (San Diego's economy being 11 times greater than Tijuana's) \cite{Bae2005}. The main source of interaction between these cities is transnational migration (i.e., people who live in one country and work, on a daily basis, in the other), even though only half the population of Tijuana has legal rights to cross the border \cite{Alegria2012}, and even less have legal right to work on the U.S. \cite{Alegria2012,Bae2005}. Nearly all of the immigrants live in Tijuana and work just across the border \cite{Bae2005}. According to Alegría \cite{Alegria2012}, 8\% of San Diego's total workforce were immigrants of Tijuana; almost all male in their thirties with secondary education. Roughly all of the migration goes through San Ysidro Land Port of Entry (LPOE), which opens 24 hours a day, seven days a week making it one of the most transited land-port in North America. It currently processes approximately 50,000 vehicles and 26,000 pedestrians per day, making it bottleneck in the system of interchange between the two countries, increasingly restricting the movement of passenger vehicles during peak times \cite{gsaYsidroplansc1}. During the day, commuters crossing San Ysidro LPOE are either going to work or returning from it; late night weekends, a population of young adult northerners are returning from the bar and nightclub districts of Tijuana \cite{Lange1999a}.

From a sociological perspective, San Diego -- Tijuana urban context has been explained as twin cities \cite{Kearney1995,viswanathan1996levy} or bi-national spaces \cite{Ganster1992,Gildersleeve1978,Sparrow2001}. Lawrence Herzog \cite{Herzog1991a,Herzog1991} has proposed the concept of \emph{trans-border} metropolis, a functional metropolitan area that physically transcend international borders, and where urban management in such areas can only be achieved through a combination of city planning and international diplomacy. Moreover, the concept of \emph{trans-border} appears to be generalizable to all border cities pairs \cite{Alegria2012}. Nonetheless, these results typically rely on data obtained from surveys and small group observation, an approach that is generally both time-consuming and expensive. On the other hand, the pervasiveness of mobile phones and location-aware social networks (such as Twitter or Foursquare) has introduced new ways to understand human mobility patterns at a national or global scale \cite{Amini2013,Blondel2011,Bora2013,Hawelka2014,Ratti2010}.

In this work, we leverage geo-located information obtained from Twitter to provide an empirical basis in which to test for the existence of a sociological construct. Using geo-located information from Twitter had been previously explored in the CSCW community: in order to understand the restrictions in communications between different communities around the globe \cite{Garcia-Gavilanes2014}, study how vulnerable users communicate during crisis \cite{DBLP:conf/cscw/KoganPA15, Olteanu2015}, or classifying Twitter users according to their information production and consumption \cite{DeChoudhury2012}. However, to the best of our knowledge, no work has used geographical information to explain transnational migrations and transnational communities.

Specifically, we focus on the following research questions:
\begin{enumerate}
  \item Within the context of geo-located social-networks, is there any evidence that Tijuana and San Diego act like a \emph{trans-border} metropolis?
  \item How do the international borders affect the mobility of the U.S. -- Mexico border metropolitan region?
\end{enumerate}

This work offers two main contributions: (i) proposing an improved approach (similar to Cranshaw's \cite{DBLP:conf/icwsm/CranshawSHS12}) to translate social-network information into a graph-theoretic framework, and (ii) providing an analysis, based on graph theory, that is capable of generating additional insights into the transnational migration and transnational community structure in the U.S. -- Mexico border region. These method and results provide a promising step for the field of Social Computing\footnote{As defined by Schuler \cite{schuler1994social}, Social Computing refers to ``any type of computing application in which software serves as an intermediary or a focus for a social relation''.}, specifically in examining online socio-behavioral phenomena. Our analysis and the conclusions drawn from it present decision-makers with a cost-effective and time saving way of sensing a transnational urban environment for which to inform the design of public policy and international diplomacy.

In the following sections we survey previous works on the context of San Diego -- Tijuana region from a sociological perspective, and works regarding exploration of urban dynamics using social network information. We also explain our method to transform geo-located social network data into a mathematical representation. Following, we present our graph analysis and community division for San Diego -- Tijuana region mobility, and a simple validation of our results based on official reports of the state of the San Diegan infrastructure. Finally, we draw our conclusions and outline possible future work. 

\section{Related Work}
\subsection{Social studies of the border}
The relation between border cities pairs has been thoroughly studied from a sociological perspective. For example, Herzog \cite{Herzog1991a,Herzog1991} argues that demographic explosions on the border have given rise to functional metropolitan areas that physically transcend international borders. This new urban environment comes with a new set of problems not merely confined to a single nation, but across international boundaries. Furthermore, Anderson and O'Dowd \cite{Anderson1999} claim that borders are no longer peripheral regions in relation to national centers, but are instead potential poles of economic growth. Border regions thus gain some independence from their national capitals when it comes to policy, and become more likely to work with their neighbors across the border in developing economic, institutional, and public infrastructure. 

Viswanathan et al.~\cite{viswanathan1996levy} examine the patterns of social distribution using 16 key variables from U.S. and Mexican Census. Their results found out that San Diego's urban planning, unlike most cities in the U.S., does not resolve around a central business district. Instead, it shows a high level of commonality with Tijuana, due to the mixing and change factors caused by nearness to the border. 

Rubin-Kurtzman et al.~\cite{Rubin-Kurtzman1996} study social and economical impacts of the southern California trans-border urban system. Their results show that the trans-border economic and technological disparities, as well as the mobility of labor and capital between southern California and Baja California affect population flows and the composition of the labor force in both sides of the border. Furthermore, the authors state that \emph{``[m]igration and trans-border mobility are the keys to demographic behavior in the region because migration is the principal determinant of population growth in Southern California and Baja California region''}.

Sparrow \cite{Sparrow2001} discusses the territorial integration of San Diego and Tijuana. On the physical level, multiple roads and highways closely link these cities. Thus, from the air, this region can be considered as one continuous urban agglomeration. On a behavioral level, there is a significant amount of work, shopping, social and touristic integration of the populations. However, there is minimal integration on the communications level, only limited cultural exchanges, and there is little to no bi-national integration on a politico-administrative level. Sparrow also argues that San Diego and Tijuana are still a long way to go as to be considered bi-national cities.

\subsection{Social Networks as Urban Sensors}
On the topic of using social networks as urban sensors, several works have shown that it is possible to consider social media participants as virtual proxies for human behavior and mobility. For example, many works have used millions of \emph{check-in}s posted in Foursquare, a social-network site that allows users to share their locations and places they visit with a group of friends. This geo-located data has been used to explain recurring patterns in human mobility \cite{cheng2011exploring}, to understand the underlying social aspects of mobility \cite{nguyen2012using}, to study the relation between distance and strength of social tie \cite{scellato2011socio}, or to identify groups of people and the places they go \cite{joseph2012beyond}. 

Cranshaw et al.~\cite{DBLP:conf/icwsm/CranshawSHS12} propose a novel approach to visualize and investigate the dynamics, structure, and character of a city on a large scale. Their method clusters Foursquare data from the city of Pittsburgh, PA into \emph{Livehoods}, an urban division that considers both the type of place and the people living and working within. This result in dynamic urban divisions, that change with people's behaviors, and are not dependent on politics or arbitrarily set divisions. Our work differs from Cranshaw's in two ways. First, although \emph{Livehoods} can be presented as an undirected weighted graph, their results do not rely on any graph analysis. In this work, we employ graph centrality measurements to further explore the relationships between neighborhoods. Secondly, whereas they study mobility from a place-centered perspective, analyzing venues and the users moving across them, we study the same phenomena from a user-centered perspective, by analyzing user movement across different places. However, we believe this difference only affects the possible interpretation of the results and not the results themselves. 

%For example, Cheng et al.~\cite{cheng2011exploring} analyze millions of check-ins posted from several locations (e.g., Foursquare). Their results show that users tend to follow simple and reproducible patterns, and that mobility is related to social status, and geographic and economic factors. Nguyen and Szymanski \cite{nguyen2012using} used a location-based social network to create and validate models of human mobility and relationships. The authors propose a friendship-based mobility model that takes into account social links. This model allows the authors to understand how often people travel together. Scellato et al.~\cite{scellato2011socio} show that among users of Brightkite, Foursquare, and Gowalla communities, 40\% of social links happen below 100 km, and there is a strong heterogeneity between users related to both social and spatial factors. Joseph et al.~\cite{joseph2012beyond} analyzed a Foursquare dataset to identify groups of people and the places they go. Their model was able to detect people who represent both geo-spatially close groups and people who appear to have similar interests.

Due to its popularity, and data access through a public API, Twitter has become one of the major sources in several works on human mobility and event detection. Some of the surveyed works have focused on the geo-tagged data representativeness and interaction with the underlying mobility dynamics \cite{quercia2012social,Jurdak2014}, while others seek to understand the relationships between geographical regions, neighborhoods and criminal behaviors \cite{Bora2013}. 

% Quercia et al.~\cite{quercia2012social} investigate how Twitter communities resemble their real-life counterparts. The authors found that most users have geographically local networks, and opinion leaders express not only positive but also negative emotions.

% Jurdak et al.~\cite{Jurdak2014} study how representative is Twitter-based mobility patterns of population level movement. Their results found out that patterns extracted from geo-tagged tweets have similar overall features as observed in mobile phone records. However clear differences arise. First, the higher resolution of Twitter data uncovers two different modes of movement: metropolitan movement and inter-city movement. Secondly, Twitter reveals unexpected dynamics in mobility for long distance movers.

% Bora et al.~\cite{Bora2013} use over ten million geo-tagged tweets from the city of Los Angeles as observations of human movement. They seek to understand the relationships of geographical regions, neighborhoods and gang territories. Their approach was to use a graph-based representation of street gang territories as nodes and interactions between them as edges. They were able to train a machine learning classifier to tell rival from non-rival links. Their results show that the distribution of the distance from home location of all tweeting activity showed heavy-tailed nature following a power law distribution (i.e., a relationship between two quantities were one varies as the power of the other).

Hawelkaa et al.~\cite{Hawelka2014} attempt to validate the representativeness of geo-located Twitter data as a global source for mobility data. Their work seeks to discover spatial patterns and clusters of regional mobility using a year of captured tweets (944 million geo-located messages). Each user is assigned to the country where he posted the most tweets, and is considered mobile if he or she issued a tweet in another country within the year. The authors build a directional country-to-country network of human travels, which enabled them to quantify both the inflow and outflow of visitors. Following the work of Sobolevsky and Newman \cite{Sobolevsky2014,newman2006modularity}, the network was split into communities or modules. The results show that travel connections between North and South America were stronger than those between America and Europe. Moreover, they were able to detect communities inside North America, South America, West and East Europe.

Analyzing information from social networks has been a recurring topic in CSCW. For example, in CSCW 2004, Goecks and Mynatt \cite{Goecks2004} presented Saori, a computational infrastructure that leverages social networks to mediate information dissemination, allowing users to share semi-public information (such as work products or their geographical location) to a small group of people. Incidentally, their results show that social-network ties that extend across a border (e.g., organizational borders or political borders) are similar to those that exist between acquaintances or colleagues, and people in shared interest groups.

In CSCW 2014, Garcia-Gavilanes et al.~\cite{Garcia-Gavilanes2014} showed that the international Twitter communication landscape was still largely dominated by geographical, economical and socio-cultural restrictions. Their work analyzes 13 million users spread over hundreds of countries. Their results show that language barrier, cultural factors dealing with intolerance and the fear of the unfamiliar are the strongest deterrence for successful collaborative work.

In CSCW 2015, Kogan et al.~\cite{DBLP:conf/cscw/KoganPA15} studied how retweeting activity, reposting or forwarding a message produced by another user, by geographically vulnerable users (e.g., those affected by hurricane Sandy in 2012) differed from the general Twitter population. Their work represents retweet activity as a directed graph, with reposterers as source nodes, the original poster as source, and directed edges between source to target representing retweets. They analyze the graph structure by looking at network size and density, degree distributions, and PageRank centrality. Their results show that hubs tend to form more during the disaster than afterwards; that geographical vulnerable users have denser interconnected networks during disasters than before or after, and that they tend to retweet information with more local utility than their non-vulnerable counterparts, who are more interested in the general picture.

In this work, we extend on these results by analyzing mobility as registered in social network within the context and with the aim of understanding the phenomenon of transnational migration. We also work upon a graph where nodes represent geographical zones and edges aggregate the number of persons living in one zone and moving to another. However, since we do not study mobility at a global scale, our zones cannot be considered whole countries, and have to be defined in a more local sense. In order to do so, we leverage the home location inference cited in Bora, with the addition of a second clustering phase, as to obtain a data-driven division of the urban environments into neighborhoods of similar density. Thus, a user belongs to a neighborhood if his or her house is within the limits defined by that neighborhood.

\section{Data and Methodology}
We used a collection of 10,908,817 geo-located tweets. Each tweet had a unique identifier, date-time of submission, coordinates of submission (i.e., GPS latitude and longitude as reported by smartphones), and the content of the message. Tweets were captured using Twitter's Streaming API\footnote{\url{https://dev.twitter.com/}} starting from December 4th, 2013 until January 13, 2015. A bounding box (32º 25' 4.2414" N, 117° 18' 49.5066" W and 33º 5' 53.3178" N, 116º 49' 17.9142" W) was used to filter only to those messages originating from San Diego and Tijuana. There are inherent limitations to our collection of data. The Twitter API provides access to approximately 1\% of all the tweets \cite{Olteanu2015}. However, \cite{DBLP:journals/corr/MorstatterPLC13} shows that the data obtained from the API closely resembles a random sample drawn from the full Twitter stream. 

After discarding all information except the tweet coordinates and the user identifier, the procedure is as follows: 
\begin{enumerate}
	\item Infer each user's home location.
	\item Cluster home locations into neighborhoods of similar density.
	\item Create a graphical representation of the neighborhoods and users moving within.
\end{enumerate}

\subsection{Inference of a user's home}
Previous work shows that it is possible to infer a user's home location using their Twitter historical data. For example, Hecht et al.~\cite{DBLP:conf/chi/HechtHSC11} performed an in-depth study of user behavior with regard of the user location field in Twitter. They found that although a huge percentage of users did not specify their location beyond a city level scale, it was possible to infer their state and city using machine-learning techniques. In another example, Bora et al.~\cite{Bora2013} assume that users are more likely to tweet from their home at night. Their method starts by filtering tweets between 7:00pm and 4:00am, then it applies a single pass of DBSCAN clustering algorithm \cite{Ester1996}. The center of the largest cluster is used as the exact coordinates for the user's home.    

This work follows up on Bora's assumption, with a minor modification: filtering tweets captured from 10:00pm to 4:00am. This new time range represents a better overlap between the American and the Mexican working cycles, since the Mexican working cycle typically extends much later than their American counterpart \cite{Booth2011}.

% \newpage
\subsection{Obtaining a neighborhood division}
Just like Cranshaw's \emph{Livehoods}, we obtain a neighborhood division by applying an additional DBSCAN into our home location data. By using DBSCAN, the resulting divisions are roughly of the same population density. This approach to producing data-driven separations of urban environments goes beyond externally imposed boundaries, such as political borough divisions or grid separations, which generally are based on census tracts and geographic landmarks \cite{DBLP:conf/icwsm/CranshawSHS12, martinezInpress}. Once again, we use the center of each cluster as the exact coordinates for each neighborhood's location. 

\subsection{Graphical approach}
A user is assigned to the neighborhood whose center is the closest to the user's home location; a tweet is assigned to the closest neighborhood by comparing the geographic coordinates reported by Twitter to each of the neighborhood's centers. Regardless of their time of submission, all tweets were considered at this stage. All distances were calculated using the Haversine formula for great-circle distances \cite{sinnott1984sky}.

We construct a directed graph depicting user's mobility. The nodes in this graph represent neighborhoods while the edges weight the number of people moving between vicinities. In other words, an edge from node $x$ to node $y$ has a weight $w$, if there are $w$ persons whose home location lies within the limits of neighborhood $x$, and tweeted at least once in location $y$.

\subsection{Experimental Settings}
The present methodology introduces a way to obtain a graph representation from a collection of geo-tagged messages, such as tweets. However, this method greatly depends on the election of the DBSCAN parameters (i.e., $\epsilon$ and \emph{minPts}) both for the clustering of individual's home locations and for the clustering of home locations into neighborhoods.

Previous work uses a \emph{minPts} parameter between 3 and 5 \cite{Ester1996,Bora2013,scikit-learn}. For the sake of considering as many users as possible, we decided on using the minimum bound for both clustering procedures. 

Our current approach does not select a single pair of $\epsilon$'s, instead it relies on a Monte Carlo simulation of a diversity of possible parameters. We produced five hundred pairs of $\epsilon_1$ and $\epsilon_2$ with the following \emph{a priori} distributions:

\begin{align*}
	\epsilon_1 &\sim U(0.3, 1.0)\\
	\epsilon_2 &\sim U(0, 0.15)
\end{align*}

where $U (a, b)$ denotes a continuous uniform distribution between $a$ and $b$. A uniform distribution was selected because it is the simplest distribution that conveys no previous knowledge of the underlying parameter distribution. However, certain assumptions are made in order to simplify the results interpretation.

The first assumption states that $\epsilon_1$ must be greater than $\epsilon_2$, based on the notion that the first cluster must coarsely capture the mobility of an individual across the span of both cities, while the second must finely distinguish between geographically close neighborhoods. Thus, the distribution for $\epsilon_1$ was selected with a lower bound twice the upper bound of $\epsilon_2$.

The second assumption states that since the distance between San Diego's north-most position and Tijuana's south-most position is less than 112km\footnote{Measured using Google Maps public API: \url{https://developers.google.com/maps/}}, all human mobility between these cities must be physically restricted to this measurement. Furthermore, considering an Earth's arc length along the equator equal to 113km\footnote{Obtained by dividing the Earth's circumference (40,075.017 km) by 360º }, then the maximum distance between San Diego and Tijuana is roughly equal to 1º. The selected upper bound for the distribution of $\epsilon_1$ reflects this assumption.

Each ordered pair ($\epsilon_1$, $\epsilon_2$) results in a different way of assigning home locations to users and different neighborhood areas. Thus, for each user there are 500 possible home locations, and 500 ways of splitting the space into neighborhoods. Unfortunately, this approach becomes computational prohibitive when considering that there would have been over two trillion ($2 \times 10^{12}$) different associations. 

In order to reduce our problem space, we propose the use of an information-based distance metric as to select only those data splits that provide the most informative separations.

\subsection{Reducing problem space with an information-based distance}
We draw a random sample of $109,088$ tweets (1\% of the dataset) and applied our two clustering procedures to obtain the 500 neighborhood divisions for this particular sample. Then, we compared all divisions pairwise using variance of information, an information-based distance metric.

Variance of information (VI) \cite{Meila2007} measures the amount of information lost or gained in changing from clustering $\mathcal{C}$ to clustering $\mathcal{C}'$ (where a clustering is a set of clusters). The algorithm takes an information-based approach, by establishing how much information is there in each clustering, and how much information one clustering gives about the other. Formally, let $\mathcal{C}$ and $\mathcal{C}'$ denote clusterings such that $\mathcal{C} = \{C_1, C_2, ..., C_K\}$ and $\mathcal{C}' = \{C'_1, C'_2, \ldots, C'_K\}$. Also, let $n_k$ denote the size of cluster $C_k$, and let $n$ be the size of the whole dataset $D$.

Assume that the probability of a point in dataset $D$ to be in cluster $C_k$ equals to 
\[
	P(k) = \frac{n_k}{n}
\]
Then a clustering's information entropy can be obtained as 

\begin{equation}
	H(\mathcal{C}) = - \sum_{k = 1}^{K} P(k) \log P(k)
	\label{eq:entropy}
\end{equation}

And the information $\mathcal{C}$ and $\mathcal{C}'$ share as 
\begin{equation}
	I(\mathcal{C}, \mathcal{C'}) = \sum_{k = 1}^{K} \sum_{k' = 1}^{K'} P(k, k') \log \frac{P(k, k')}{P(k) P(k')}
	\label{eq:mutualinfo}
\end{equation}
where $P(k,k')$ represents the joint probability distribution of the variables associated with the clusterings, that is, the probability that a point belongs to $C_k$ in clustering $\mathcal{C}$ and to $C'_k$ in $\mathcal{C'}$. Finally, VI can be defined using equation \eqref{eq:entropy} and \eqref{eq:mutualinfo}, as 

\begin{equation}
	VI(\mathcal{C}, \mathcal{C'}) = H(\mathcal{C}) + H(\mathcal{C'}) - 2I(\mathcal{C}, \mathcal{C'})
	\label{eq:VI}
\end{equation}

This process outputs a distance distribution over the information contained in each division. For this work, only the top 5\% most informative comparisons were considered. Between them, the top $5$ percentile clusterings divided the Tijuana-San Diego region into $5,628$ areas. Finally, our approach considers all this areas as part of a directed graph. Just as mentioned earlier, this areas correspond to graph nodes and the edges sum up the number of people moving between them.

\section{Results}
\begin{figure}
	\begin{center}
		\includegraphics[width=\columnwidth]{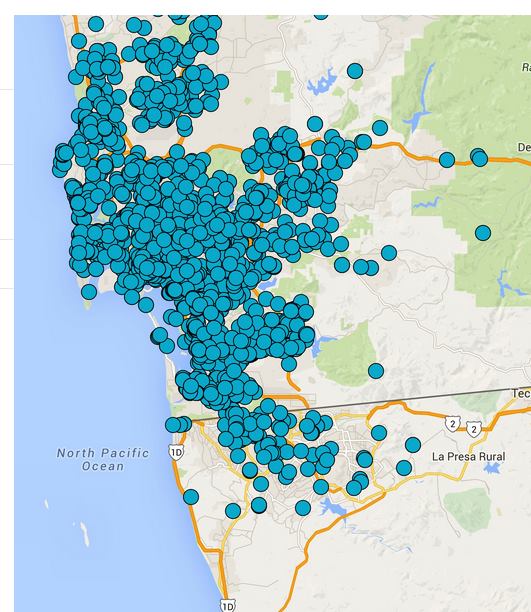}
	\end{center}
	\caption{Spatial embedding of the mobility graph for Tijuana and San Diego. 5,576 vertices formed the graph. For the sake of clarity, edges are omitted.  }
	\label{fig:TJ_SD_mobGraph}
\end{figure}

The mobility between San Diego and Tijuana can be represented as a directed graph formed by $5,628$ nodes with $212,572$ connections between them. Thus, the graph was not completely connected. Instead it was formed by $53$ different pieces (components). All nodes not connected to any other node were eliminated, leaving the final graph with $5,576$ vertices and $211,796$ edges. Figure \ref{fig:TJ_SD_mobGraph} shows the spatial embedding of the mobility graph.

\begin{figure}
	\begin{center}
		\includegraphics[width=\columnwidth]{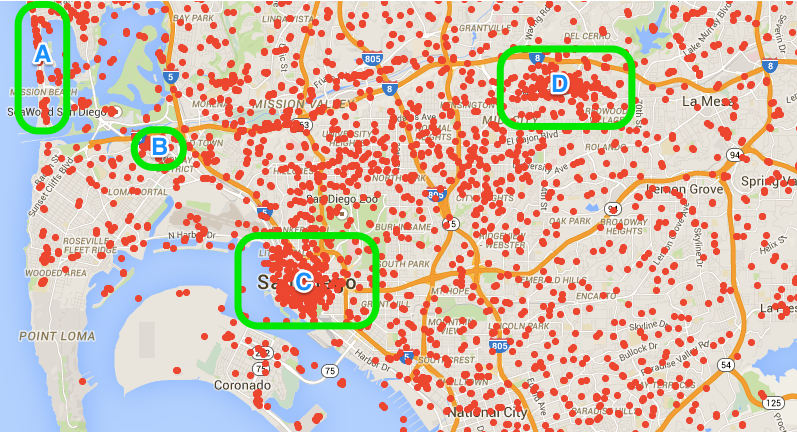}
	\end{center}
	\caption{Concentration of nodes reveals important areas of San Diego: (A) Mission Beach, a popular recreational area and home to attractions such as SeaWorld; (B) Valley View Casino Center, the city's main sports arena; (C) Downtown San Diego, and (D) UCSD. }
	\label{fig:interestingZones}
\end{figure}

\begin{figure}
	\begin{center}
		\includegraphics[width=\columnwidth]{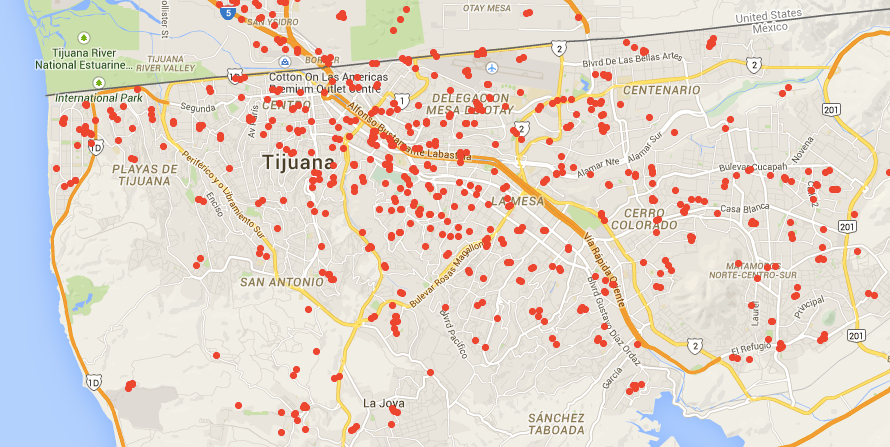}
	\end{center}
	\caption{In the case of Tijuana, nodes were scarce and pretty evenly spaced, making the visual recognition of clusters difficult.}
	\label{fig:notinterestingZones}
\end{figure}

Concentration of nodes in certain areas allowed for the identification of some important San Diegan locations. For example, University of California at San Diego (UCSD), the city's downtown, Valley View Casino Center (formerly San Diego Sports Arena), and Mission Beach. Figure \ref{fig:interestingZones} highlights the main areas. On the Mexican side, this analysis was not as successful. Nodes in the Tijuana area were scarce and fairly spread out, making concentrations of no more than 3 or 4 points at the time (see Figure \ref{fig:notinterestingZones}).

\begin{figure*}
  \centering
  \includegraphics[width=2\columnwidth]{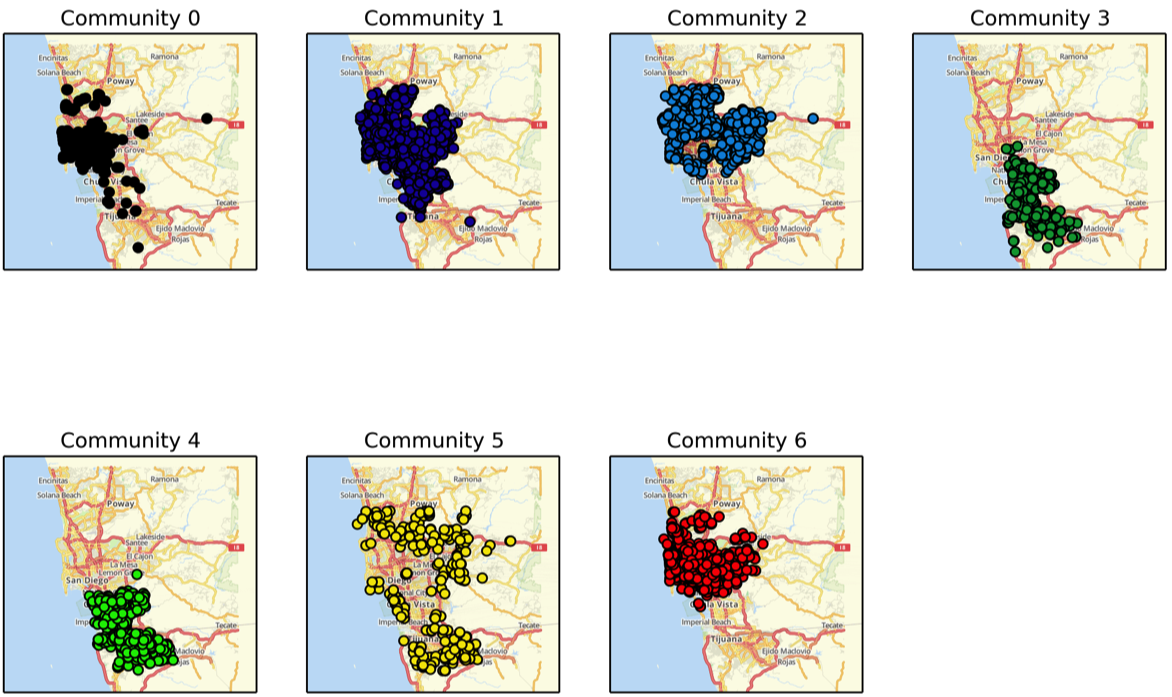}
  \caption{Communities of Tijuana and San Diego. Mexican side of the border is deeply connected to the San Diegan side; however, there are communities in San Diego whose movements are completely restricted to the U.S. territory.}
  \label{fig:TJ_SD_communities}
\end{figure*}

% \newpage
\subsection{Graph Analysis}
An analysis of node degree and \emph{betweenness} centrality allows for the determination of key points for understanding trans-border mobility. Moreover, the fact that the graph was not completely connected permits an analysis of the graph's modularity, a measure of how many communities can the graph be split into, and hence a study of the transnational migration patterns. 

\subsubsection{Node degree}
Weighted in-degree (i.e., the number of people arriving at each node) and weighted out-degree (i.e., the number of people leaving each node) were obtained. Weighted In-degree had a distribution with minimum $1$, maximum $58,195$, and median $115$. Out-Degree ranged from $1$ to $62433$ with a median of $112$. There was no statistical difference between the two distributions (KS-test $KS = 0.0204,\, p > 0.05$). The node with the highest degree in-bound was located in the north-most part of Coronado peninsula, corresponding to Naval Air Station North Island, and the node with highest degree out-bound was 32º 41' 7.8" N 117º 02' 22.0" W, which does not directly point to any landmark. We hypothesize that such a high out-degree was due to the proximity of California State Road 54 (SR 54) which connects Interstate 5 (I-5) and El Cajon, California.

\subsubsection{Communities and modularity}
A \emph{community structure} is the appearance of densely connected groups of vertices that have only sparse connection with other groups \cite{newman2006modularity}. Modularity is the measurement of the division of a network into communities or modules. Commonly denoted by $Q$, it presents a real value between $-\frac{1}{2}$ (inclusive) and $1$ (exclusive) where higher values indicate the presence of \emph{community structure} within the network. Using the Louvain method \cite{blondel2008fast}, seven distinct communities were detected, with a value of $Q = 0.707$. 

Figure 4 shows the communities in relation to the border. Three out of the seven communities (community 3, 4 and 5) depict a constant mobility across U.S. -- Mexico frontier. Two other communities (community 0 and 1) have only a few nodes across the border. The two communities (2 and 6) contained all its mobility within the U.S. side of the border. No community was restricted to the Mexican side. These results hint towards three types of transnational migration: (i) a flow that continuously crosses the border, from Mexico up to Chula Vista, California and vice versa; (ii) a group of people who live in San Diego City and has no need to cross the border at all, and (iii) a group of people commuting from Tijuana all the way up to La Jolla, California. Communities that do not cross the border have a much more confided mobility than their international counterparts. Community 0 captures mobility from and to San Diego City's downtown, as well as movement from and to UCSD. San Ysidro land point of entry (LPOE), Tijuana's main border crossing, is also represented inside community 0 as one of the points over the border. Community 2 depicts movement from El Cajon, CA and northwest areas towards La Jolla, CA. Communities 3 and 4 represent the mobility in Tijuana all the way north up to National City, CA. Community 5 movement goes all the way up to La Jolla, CA, where the limits of this study were set. However, it is quite possible that human journeys from this community would had reached Los Angeles area, hadn't the limit existed. Figure 5 and Figure 6 illustrate three of the communities that go across the border in a greater detail.

\subsubsection{Node Centrality}
A node's centrality is a measure of the importance of this particular node within the network \cite{newman2010networks}. Centrality measures allow for key element identification, especially important in biology (e.g., to understand main decease spreaders) or social network analysis (e.g., to identify influential people). Many of these centralities are based on shortest paths linking pairs of nodes \cite{brandes2001faster}. One of such centralities is Betweenness Centrality (BC) \cite{anthonisse1971rush,freeman1977set}, which measures the ability of an individual node to control the communication flow in the network \cite{Altshuler2011}. BC is normally calculated as the fraction of shortest paths between node pairs that pass through the node of interest \cite{Newman200539}. Extensions to the BC definition made it applicable to weighted graphs \cite{brandes2008variants,white1994betweenness}. Furthermore, several works have shown a positive correlation between traffic congestion in a transportation network and its corresponding node's BC measurement \cite{Altshuler2011,grindrod2011communicability,holme2003congestion}

\begin{figure}
	\centering
	\includegraphics[width=0.7\columnwidth]{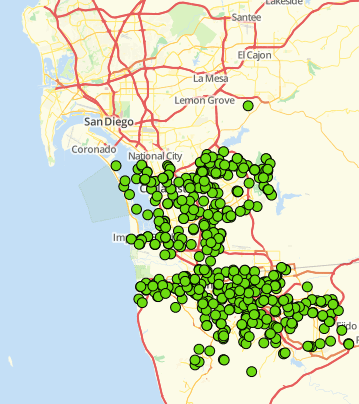}
	\caption{Community 4 spans movement from Tijuana and National City, California. The sub-graph here presented was fully connected, although the edges are omitted again for clarity. }
	\label{subfig:TJ_SD_comm4}
\end{figure}

\begin{figure}
  \centering
	\includegraphics[width=0.7\columnwidth]{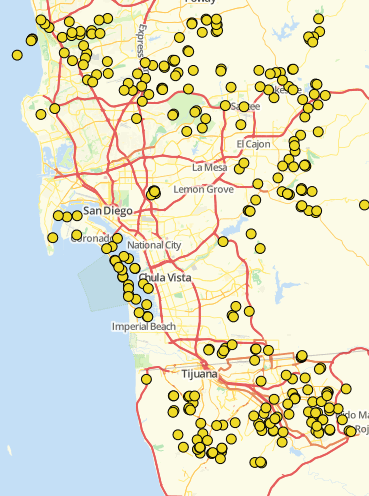}
	\caption{Community 5 that spans all the way up to La Jolla. Again, the sub-graph was fully connected. Edges are omitted for clarity.}
	\label{subfig:TJ_SD_comm5}
\end{figure}

Every node's BC was calculated. The obtained measurements lay between $0.0$ and $2,708,509.9$, with a median of $2,262.7$. This distribution was highly skewed to the left (i.e., only a few nodes had huge values while the majority had really low values). The Montogmery Field Airport found the highest value just short of the intersection between Interstate 805, the major north-south highway in Southern California, and SR163. The second highest BC was found closely to San Ysidro LPOE. The third most important BC was found in 32º 41' 7.8" N 117º 02' 22.0" W (the same node that scored the highest out-degree earlier on) between SR 54, Woodman Street and Briarwood road. Once again we believe this result was due to the importance of SR 54 in connecting I-5 and El Cajon, CA. On fourth place, there is a node on the Navy's Lodge in Coronaro peninsula. Finally, the fifth node in importance is located in Terra Nova Chula Vista, an apartment complex close to I-805. We believe this is because this node is intended to capture journeys traveling on the I-805.

\subsection{Validation of the results}
Just like Hawelkawa et al.~\cite{Hawelka2014} mention, it is difficult to find a bias-free human mobility dataset that would enable direct validation of results obtained with Twitter. A remaining possibility would be to use existing traffic services such as Google Maps or Waze. The latter even has its own public API for traffic\footnote{\url{https: //www.waze.com/about/dev}} . However, obtaining information from these services would require a development that goes beyond the scope of this work. Instead, we draw a comparison of our results and those presented by official government agencies reporting on the status of San Diego's infrastructure.

The most central node was found on SR 163, close to the intersection of I-805. Interstate 805 (I-805) is a major north/south freeway whose primary purpose is to provide an alternative route for I-5 traffic in order to bypass the congested Central Business District (CBD) \cite{californiaDistrict11-805}. It also serves as a commuter route providing direct access to employment centers in Otay Mesa, Kearny Mesa and Sorrento Valley. Along with I-5, I-805 is an important corridor for the movement of people and goods from Baja California and the U.S. -- Mexico border region to the northern destinations \cite{californiaDistrict11-805}. The exits from I-805 towards SR-163, as well as the entry from SR-163 towards I-805, were found to be operating at a highly deficient level of service \cite{californiaDistrict11-805}. In 2008, it served almost 200,000 commuters on average every weekday. It is estimated to have over 250,000 projected average weekday daily traffic by 2030 \cite{californiaDistrict11-163,californiaDistrict11-805}.

The second most important node was just short of the San Ysidro border crossing. San Ysidro border crossing is one of the busiest land border crossings in the world. Open 24 hours; 7 days a week it handles a daily traffic of 50,000 vehicles and 26,000 pedestrians per day \cite{dibbleBorder,gsaYsidroplansc1}. A 2005 study from the San Diego Association of Governments, in cooperation with the California Department of Transportation, found that San Diego lost over \$1.3 billion in potential revenues; 3 million working hours; and 28,000 to 35,000 jobs because of excessive border waits \cite{sandag05}. A recent expansion project from the U.S. plummeted San Ysidro border crossing wait times to just minutes \cite{KBPS}.

% \balance{}
According to BC, the third most important node was found between SR 54, and Woordman St. SR 54 is a major east-west facility serving intraregional traffic, providing access to the communities in the South Bay, Spring Valley, Rancho San Diego, and the cities of Chula Vista, Nacional City and El Cajon. SR-54 provides an alternative route to I-805, SR-94 and I-8. Travelers to Mexico can reach I-5, I-805, SR-194 and SR-125 by way of SR-54. In 2009, SR-54 between I-5 and I-805, and between I-805 to Brianwoord scored a D-rank level of service, with close to 126,000 and 118,000 weekday average number of commuters. By 2030, this road is expected to serve almost 150,000 people on an average weekday \cite{californiaDistrict11-54}.

\section{Discussion}
We obtained 10.9 million tweets from San Diego -- Tijuana border region. The proposed method infers the home location of the users and their mobility through the region. By applying again a clustering procedure to the home locations, we were able to divide the urban space into neighborhoods of similar density. This neighborhoods are data-driven, thus free from constrains of current political divisions (e.g., they do not have to follow streets or landmarks). We represent the region's mobility as a directed graph by using neighborhoods as nodes, and assigning weights to the edges according to the number of people living in one location and traveling to another. We shall now discuss the main implications of our findings, main limitations, and possible future works. 

\subsection{Implications}
\textbf{Within the context of geo-located social-networks, is there any evidence that Tijuana and San Diego act like a trans-border metropolis?} We obtained the community structure of San Diego County and Tijuana by using a community detection algorithm on the mobility graph. We found seven communities that explain the region's human movement. People who live on one side of the border and constantly cross over to the other side formed five of these. We believe most of this movement is due to transnational migration, since no community restricted its movement to the Mexican side. On the other hand, some of the groups moved only on the San Diego City region, in a pattern that seemed specifically to people living and commuting close to the city's business district. These results support the idea of a trans-border region closely linked by infrastructure and by daily commuters crossing the border for their economical and leisure activities. Our results support Sparrow's and Alegría's \cite{Alegria2012,Sparrow2001} claims that, on a infrastructural and behavioral level, these cities are closely linked. However, at the current time, we cannot support nor disclaim Sparrow's asseveration on the lack of cultural and politico-administrative integration of the region \cite{Sparrow2001}. However, García-Gavilanes \cite{Garcia-Gavilanes2014} already showed that trans-border communication in Twitter is highly limited by language barriers and differences in cultural factors, both of which are present in San Diego -- Tijuana region. 

% \newpage
\textbf{How do the international borders affect the mobility of the U.S. -- Mexico border metropolitan region?} Mobility between San Diego and Tijuana is largely dictated by their land ports of entry (LPOE), specially regarding the main border crossing in San Ysidro. Previous works have already shown the importance of San Ysidro LPOE in understanding economics, migration and mobility of labor in the region. Even the U.S. General Services Administration has noted that San Ysidro LPOE has become a bottleneck in the system of interchange between the two countries, increasingly restricting the movement of passenger vehicles during peak times \cite{gsaYsidroplansc1}. Our centrality measurements confirmed this pattern. The graph's node corresponding to San Ysidro LPOE was ranked second in importance, just behind the SR 163 and I-805 intersection.

\subsection{Limitations and Extensions}
The conclusions of this paper are limited in scope by a sample that is not representative of the general population, only of Twitter users with geo-tagging enabled. Previous works have noted that geo-located tweets account for around 1\% of the total of messages submitted to Twitter \cite{Hawelka2014,DBLP:journals/corr/MorstatterPLC13}. However, future work might circumvent this limitation by feeding from distinct geo-located public data sources. This would result in a better understanding of the different behaviors of users in different modes of transportation, improve on the possible routing of human transit, and reduce the sample bias suffered from using only a Twitter sample.

Moreover, the geographical restrictions of our study were limited to Southern California and Tijuana. Additional explorations of our results hint towards mobility flow heading for Calexico, California and its sister city Mexicali, Baja California. Future work might be able to find that some communities' mobility extend beyond the San Diego metropolitan area.

\section{Conclusions}
In this work, we have studied geo-located social-network information obtained from Twitter to provide an empirical basis to further understand the trans-border metropolis and their transnational migration dynamics. Our aim was to understand the daily commutes of people living in the San Diego -- Tijuana transnational metropolitan region. We have employed a methodology capable of translating social-network information into graphical representation. This graph and the analysis that can be performed on it have considerable potential value to policy makers and urban planners. Our methodology and results intent to provide urban-planners and decision-makers with a time-saving, easy-to-deploy, and cost-effective method of sensing an urban environment as to assert population's mobility and inform the design of public policy, particularly with respect to transportation and immigration topics. This is especially important when considering that urban planning in a future trans-border megalopolis will require knowledge of multiple (and highly distinct) cultural, social and economical factors.

% \section{Acknowledgments}
% [TODO]

% \newpage
% REFERENCES FORMAT
% References must be the same font size as other body text.
\bibliographystyle{SIGCHI-Reference-Format}
\bibliography{referencias}

\end{document}